\documentclass[12pt]{article}

\usepackage[a4paper,left=30mm,right=20mm,top=20mm,bottom=20mm]{geometry}
\usepackage{graphics}
\usepackage{amsmath}
\usepackage{amssymb}
\usepackage{epsfig}
\usepackage{cite}
\usepackage{xcolor}
\usepackage[unicode]{hyperref}
\newcommand{\be}{\begin{eqnarray}}
\newcommand{\ee}{\end{eqnarray}}
\newcommand{\beq}{\begin{equation}}
\newcommand{\eeq}{\end{equation}}

\title{Refined TMD gluon density in a proton from the HERA and LHC data}

\author{A.V.~Lipatov$^{1,2}$, G.I.~Lykasov$^{2}$, M.A.~Malyshev$^{1,3}$}

\begin{document}

\maketitle

\begin{center}

{\it $^{1}$Skobeltsyn Institute of Nuclear Physics, Lomonosov Moscow State University, 119991, Moscow, Russia}\\
{\it $^{2}$Joint Institute for Nuclear Research, 141980, Dubna, Moscow region, Russia}\\
{\it $^{3}$Moscow Aviation Institute, 125993, Moscow,  Russia}\\

\end{center}

\vspace{0.5cm}

\begin{center}

{\bf Abstract }

\end{center}

\indent 

We update the phenomenological parameters of the Transverse Momentum Dependent (TMD, or unintegrated) 
gluon density in a proton proposed in our previous studies.
This analysis is based on the analytical expression for starting gluon
distribution which provides a self-consistent simultaneous description of
HERA data on proton structure function $F_2(x,Q^2)$, reduced cross section for the 
electron-proton deep inelastic scattering at low $Q^2$ 
and soft hadron production in $pp$ collisions at the LHC conditions.
We extend it to the whole kinematical region using the 
Catani-Ciafaloni-Fiorani-Marchesini (CCFM) evolution equation.
Exploiting our previous results on a number of semihard QCD processes, we performed a combined fit 
to an extended set of LHC and HERA data, comprising a total of $509$ points from $16$ data sets.
We illustrate 
our fit by applying
the derived TMD gluon density in a proton to inclusive prompt photon photoproduction at HERA.

\vspace{1cm}


\newpage

It is well known that parton distribution functions in a proton (PDFs), $f_a(x,\mu^2)$ with $a = q$ or $g$, are an essential 
ingredient of any description of hard scattering at modern colliders energies. 
If only one scale is present in the process, $\mu\sim\sqrt{s}\gg\Lambda_\text{QCD}$, 
then the PDFs can be described in Quantum Chromodynamics (QCD) via the Dokshitzer-Gribov-Lipatov-Altarelli-Parisi (DGLAP) 
equations\cite{DGLAP}. 
However, in case of a two-scale process, $\sqrt{s}\gg\mu\gg\Lambda_\text{QCD}$, 
the gluon dynamics 
can be described by the Balitsky-Fadin-Kuraev-Lipatov (BFKL)\cite{BFKL} or 
Catani-Ciafaloni-Fiorani-Marchesini (CCFM)\cite{CCFM} equations. 
It leads to Transverse Momentum Dependent (TMD, or unintegrated) 
gluon densities in a proton and high energy factorization\cite{HighEnergyFactorization}, or $k_T$-factorization\cite{kt-factorization} 
approach (see review\cite{TMD-review} for more information).
The $k_T$-factorization approach is quite widely used in phenomenological applications and implemented in 
several Monte-Carlo event generators, such as \textsc{pegasus}\cite{PEGASUS} and
\textsc{cascade}\cite{CASCADE}. 
A comprehensive set of TMD gluon distributions in a proton is available in the \textsc{tmdlib} library\cite{TMDLib2}.

In general, TMD gluon densities can be calculated within some approaches, 
such as popular Kimber-Martin-Ryskin formalism\cite{KMR-LO, KMR-NLO}, Parton Branching approach\cite{PB1, PB2} or obtained from the analytical
or numerical solutions of BFKL-like QCD evolution equations\footnote{There are also investigations 
within the non-linear evolution in QCD (see, for example,\cite{JIMWLK-1, JIMWLK-2, JIMWLK-3, JIMWLK-4, JIMWLK-5, JIMWLK-6, BK, Unified-BK-DGLAP, Kutak-Sapeta} and references therein).}.
The CCFM equation, which resumes large logarithmic terms proportional to $\alpha_s^n \ln^n 1/x$ and
$\alpha_s^n \ln^n 1/(1 - x)$ and therefore valid at both low and large $x$,
has been applied\cite{JH2013}. 
In these calculations, the empirical expression for initial gluon density at some 
starting scale $\mu_0$ (which is of the order of the hadron scale) with 
factorized Gauss smearing in transverse momentum was applied.
In contrast, in our previous study\cite{LLM-2022} a more physically motivated expression for the
input distribution was chosen.
It is based on the description of the LHC data on soft hadron transverse momenta spectra
in the framework of the modified soft quark gluon string 
model\cite{ModifiedSoftQuarkGluonStringModel-1, ModifiedSoftQuarkGluonStringModel-2} (see also\cite{SoftQuarkGluonStringModel-1, SoftQuarkGluonStringModel-2}) with taking into 
account gluon saturation effects important at small $x$ and low scales\cite{Saturation-Mueller}.
The obtained CCFM-evolved gluon density in a proton (LLM)
was successfully tested later on a number of collider data, 
in particular, on latest HERA data on longitudinal structure function of proton\footnote{A comparison 
of LLM predictions on $F_L(x,Q^2)$ with other models can be found\cite{FL-Boroun}.} $F_L(x,Q^2)$\cite{FL-our} 
and associated photoproduction of prompt photons and jets\cite{Photon-our}.

Very recently it was shown\cite{LLM-F2} that some 
phenomenological parameters of the starting LLM gluon density
need to be corrected in order to provide a good description
(within the color dipole approach\cite{DP-NZ-1, DP-NZ-2, GBW1, GBW2}) of the low $Q^2$ data on proton structure 
function $F_2(x,Q^2)$ and reduced deep inelastic cross sections taken by H1 and ZEUS Collaborations.
These parameters are related mainly with the small $x$ behaviour of the TMD gluon
density at low scales, as it will demostrated below.
At the same time, such an adjustment does not spoil the quality of soft hadron spectra data, provided the 
corresponding parameters of fragmentation of quarks and diquarks to the hadrons are altered moderately (see\cite{LLM-F2} for more details).
Of course, in this case essential parameters, which cannot be determined from 
these data, have to be fitted from other measurements.
It is the aim of this short Letter to perform such a fit using an extended set 
of experimental data for several processes known to be sensitive to the gluon content of the proton
and then complete our study by applying the CCFM evolution to the newly fitted initial density.
As a main result, we present a more universal TMD gluon distribution function in a proton,
which is already available in the Monte-Carlo event generator \textsc{pegasus} and \textsc{tmdlib} tool.

\begin{figure}
\begin{center}
\includegraphics[width=7.3cm]{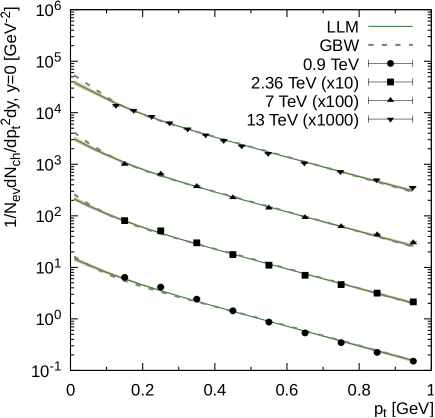} \hspace{1cm}
\includegraphics[width=7.3cm]{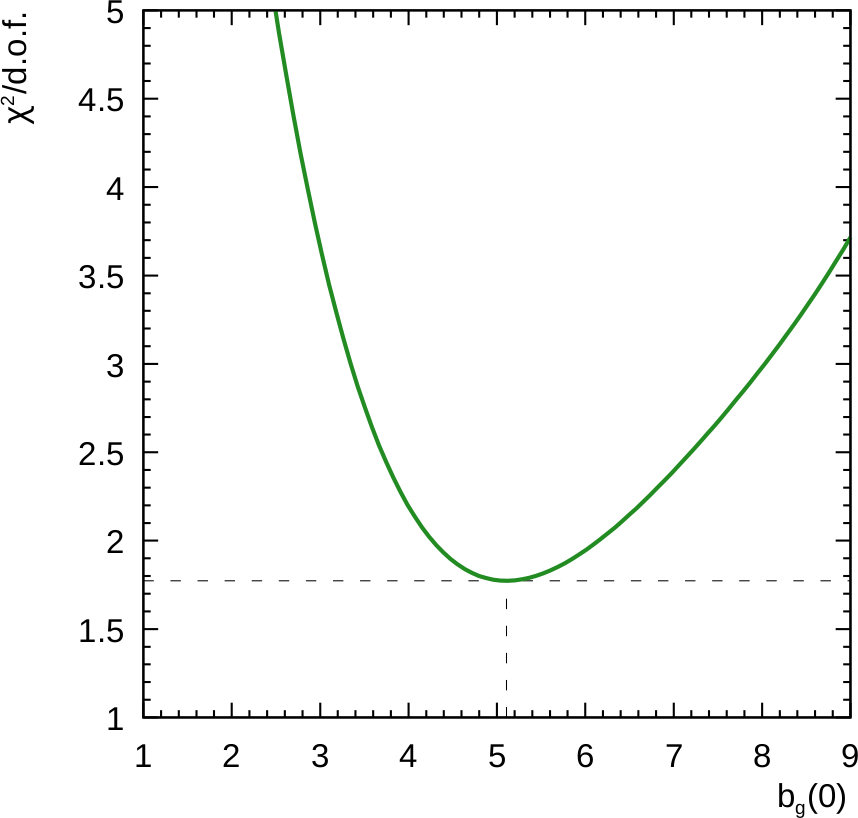}
\caption{Left panel: transverse momentum distributions of multiplicities of soft charged hadrons produced in $pp$
collisions at different LHC energies in the mid-rapidity region. 
Shaded bands correspond to the uncertainties of our calculations connected with
the uncertainties in determination of $x_0$ and parton-to-hadron fragmentation parameters.
The experimental data are from ATLAS\cite{SoftHadronData-1, SoftHadronData-3}
and CMS\cite{SoftHadronData-2}. Right panel: $\chi^2/d.o.f.$ dependence of our fit for $b_g(0)$ parameter.}
\label{fig1} 
\end{center}
\end{figure}

So, we ended up with the following form (non-factorized with respect to $x$ and ${\mathbf k}_T^2$) of the initial LLM gluon density in a proton:
\begin{gather}
  f_{g}(x, {\mathbf k}_T^2) = c_g (1-x)^{b_g} \sum_{n = 1}^3 c_n \left[R_0(x) |{\mathbf k}_{T}|\right]^n \exp \left(-R_0(x) |{\mathbf k}_{T}|\right), 
\label{eq:OurGluon1}
\end{gather}
\noindent 
where
\begin{gather}
  R_0^2(x) = {1\over Q_0^2}\left(x\over x_0\right)^\lambda, \quad b_g = b_g(0) + {4C_A\over \beta_0}\ln{\alpha_s(Q_0^2)\over \alpha_s({\mathbf k}_{T}^2)},
\label{eq:OurGluon2}
\end{gather}
\noindent
with $C_A = N_C$, $\beta_0 = 11 - 2N_f/3$ and $Q_0 = 2.2$~GeV.
Here $b_g$ parameter is treated to be running at ${\mathbf k}_T^2 > Q_0^2$ only,
whereas the fixed value $b_g = b_g(0)$ at ${\mathbf k}_T^2 \leq Q_0^2$ is used.
Simultaneous best fit to the HERA data on reduced deep inelastic cross section $\sigma_r(x,Q^2)$ and
proton structure function $F_2(x,Q^2)$ 
measured at low $Q^2 < 4.5$~GeV$^2$ 
and LHC data on charged hadron (pion and kaon) production at small transverse momenta $p_T$ 
in the mid-rapidity region leads to $c_1 = 5$, $c_2 = 3$, $c_3 = 2$, $x_0 = 1.3 \cdot 10^{-11}$ and $\lambda = 0.22$\cite{LLM-F2}.
The experimental data on charged hadron production involved into the fit are compared with our predictions in Fig.~\ref{fig1} (left panel). 
One can see that good agreement is achieved in a wide range of energies.
For a comparison we also show the results obtained within the popular GBW model\cite{GBW1,GBW2}.
It was argued\cite{LLM-F2} that the latter gives a somewhat worse description of the data.

Our procedure to determine remaining parameters, $b_g(0)$ and $c_g$, follows then the investigation\cite{LLM-2022}. 
As usual, they are extracted by
minimizing
\begin{gather}
  \chi^2 = \sum_i \left[ { {\cal C}_i^{\rm data} - {\cal C}_i^{\rm theory} \over \Delta {\cal C}_i^{\rm data} } \right]^2
\end{gather}
\noindent
where $i$ runs over all experimental data points for observables ${{\cal C}_i}$ from analyses listed in Table~\ref{tbl:DataSets}.
Exploiting our previous results,
theoretical predictions for each of ${{\cal C}_i}$ are generated for large number of fixed guessed
$b_g(0)$ values in the range $1 \leq b_g(0) \leq 9$ after applying the CCFM evolution to input density~(\ref{eq:OurGluon1}).
At this step we used \textsc{updfevolv} routine\cite{uPDFevolv} to
solve numerically the CCFM equation.
Note that the data set is extended now compared to the previous analysis\cite{LLM-2022}
by including the ATLAS data\cite{bjet-ATLAS-7} on inclusive $b$-jet production in $pp$ collisions at $\sqrt s = 7$~TeV,
CMS data on $c$-jet production in $pp$ interactions at $\sqrt s = 2.76$ and $5.02$~GeV\cite{cjet-CMS-2.76} and HERA data\cite{gamma-H1, gamma-ZEUS} on 
inclusive prompt photon photoproduction in $ep$ collisions at $\sqrt s = 319$~GeV.
Our simultaneous fit to all data sets leads to $b_g(0)=5.109$ and $c_g=1.2708$
with goodness $\chi^2/d.o.f. = 1.773$ for $509$ data points, see Fig.~\ref{fig1} (right panel).
Then, to obtain the TMD gluon density in the whole kinematical range, we solved 
the CCFM equation numerically using the fitted values of the parameters above. 
In this way, the gluon distribution function (LLM) is tabulated in a commonly recognized format (namely, grid of $50 \times 50 \times 50$ bins in $x$, ${\mathbf k}_T^2$ 
and $\mu^2$) which is used in the \textsc{tmdlib} package.
It is already available for public usage from there and implemented also into the Monte-Carlo event generator \textsc{pegasus}.

\begin{table} \footnotesize 
\label{table1}
\begin{center}
\begin{tabular}{|c|c|c|c|c|c|c|}
\hline
 Experiment        & Collaboration & Year & Reference & Collision & $\sqrt s/{\rm GeV}$ & Number of points \\
\hline
incl. $c$-jet & CMS & $2017$ & \cite{cjet-CMS-2.76}  & $pp$ & $2.76$  & $5$ \\
\hline
incl. $c$-jet & CMS & $2017$ & \cite{cjet-CMS-2.76}  & $pp$ & $5.02$  & $5$ \\
\hline
incl. $b$-jet & ATLAS & $2011$ & \cite{bjet-ATLAS-7}  & $pp$ & $7$  & $46$ \\
\hline
incl. $b$-jet & CMS & $2012$ & \cite{bjet-CMS-7}  & $pp$ & $7$  & $98$ \\
\hline
$F_2^c(x, Q^2)$ & H1 & $2010$, $2011$ & \cite{F2cb-H1, F2c-H1}  & $ep$ & $0.319$  & $25$ \\
\hline
$F_2^c(x, Q^2)$ & ZEUS & $2014$ & \cite{F2cb-ZEUS}  & $ep$ & $0.319$  & $18$ \\
\hline
$F_2^b(x, Q^2)$ & H1 & $2014$ & \cite{F2cb-H1}  & $ep$ & $0.319$  & $12$ \\
\hline
$F_2^b(x, Q^2)$ & ZEUS & $2014$ & \cite{F2cb-ZEUS}  & $ep$ & $0.319$  & $17$ \\
\hline
$\sigma_{\rm red}^c(x, Q^2)$ & H1, ZEUS & $2018$ & \cite{sigma_red-ZEUS+H1}  & $ep$ & $0.319$  & $51$ \\
\hline
$\sigma_{\rm red}^b(x, Q^2)$ & H1, ZEUS & $2018$ & \cite{sigma_red-ZEUS+H1}  & $ep$ & $0.319$  & $27$ \\
\hline
incl. $H \to \gamma \gamma$ & CMS & $2023$ & \cite{HVV-CMS}  & $pp$ & $13$  & $37$ \\
\hline
incl. $H \to \gamma \gamma$ & ATLAS & $2018$ & \cite{HVV-CMS}  & $pp$ & $13$  & $27$ \\
\hline
incl. $H \to ZZ^*$ & CMS & $2023$ & \cite{HZZ-CMS}  & $pp$ & $13$  & $54$ \\
\hline
incl. $H \to ZZ^*$ & ATLAS & $2020$ & \cite{HZZ-ATLAS}  & $pp$ & $13$  & $54$ \\
\hline
incl. $\gamma$ & H1 & $2010$ & \cite{gamma-H1}  & $ep$, low $Q^2$ & $0.319$  & $25$ \\
\hline
incl. $\gamma$ & ZEUS & $2014$ & \cite{gamma-ZEUS}  & $ep$, low $Q^2$ & $0.319$  & $8$ \\
\hline
\end{tabular}
\end{center}
\caption{The list of experimental data used for the fitting procedure and number of data points for each experiment.}
\label{tbl:DataSets}
\end{table}

\begin{figure}
\begin{center}
\includegraphics[width=7.3cm]{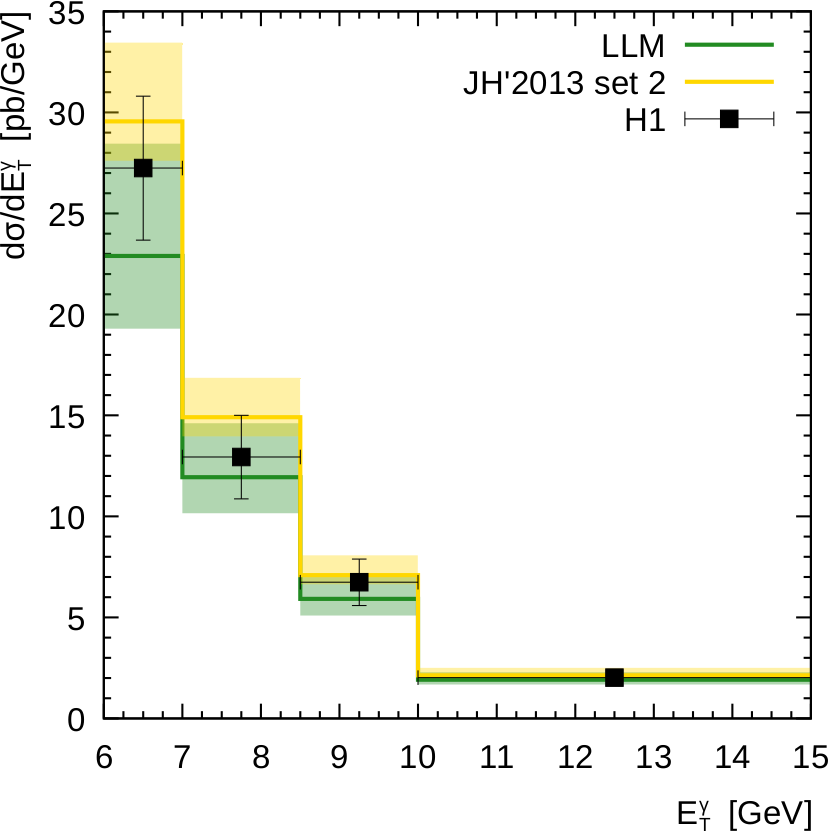} \hspace{1cm} \vspace{1cm}
\includegraphics[width=7.3cm]{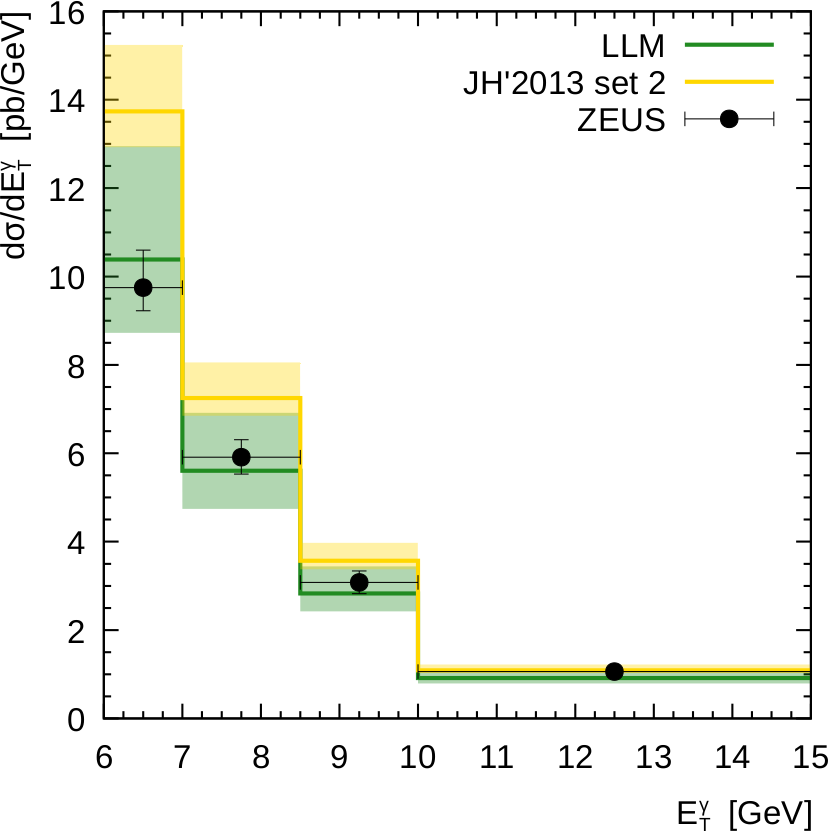} 
\includegraphics[width=7.3cm]{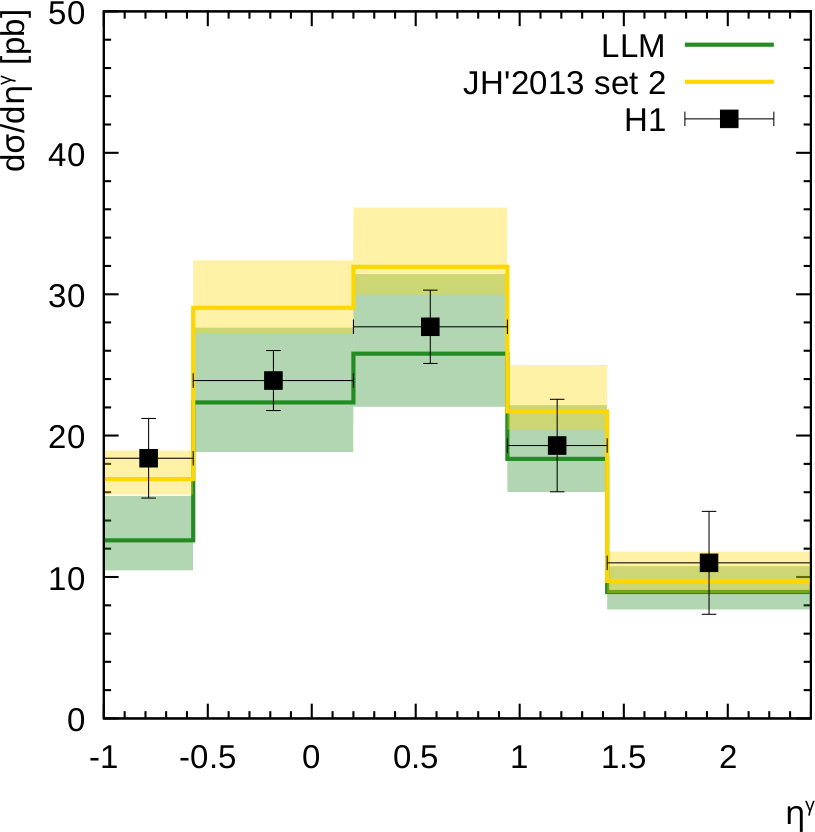} \hspace{1cm} 
\includegraphics[width=7.3cm]{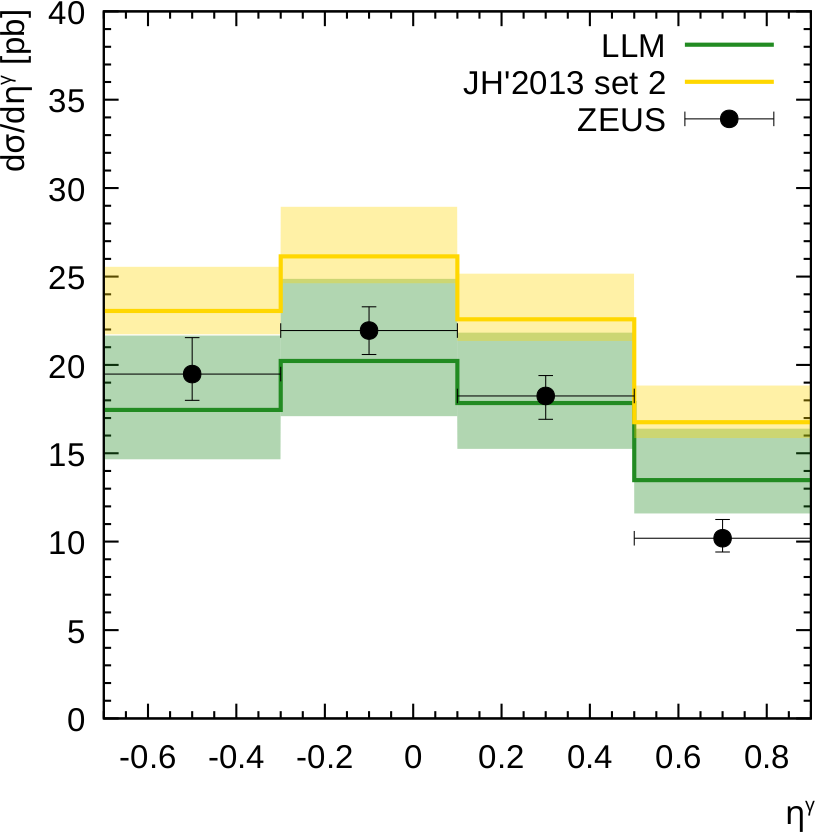}
\caption{The inclusive prompt photon photoproduction cross
section as functions of the photon transverse energy and pseudorapidity. 
The green (yellow) histograms and shaded bands correspond to the predictions obtained
with the LLM (JH'2013 set 2) gluon density and estimated scale uncertainties of these calculations.
The experimental data are from H1~\cite{PromptPhoton-H1} and ZEUS~\cite{PromptPhoton-ZEUS1}.}
\label{fig2} 
\end{center}
\end{figure}

Let us now illustrate our fit with the inclusive prompt photon photoproduction in $ep$ collisions at HERA,
where sensitivity to the gluon density in a proton can be seen clearly.
Here, in the photoproduction regime of DIS, the colliding electron emits a quasi-real photon, 
which then interacts with a proton. 
Here we completely follow our previous study\cite{Photon-our}, where we have
investigated the prompt photon photoproduction with associated hadronic jets. 
The consideration is based on two leading off-shell (depending on the non-zero 
virtualities of incoming gluons) photon-gluon subprocesses, 
namely $\gamma g^*\to \gamma q\bar q$ and 'box' subprocess $\gamma g^*\to \gamma g$, currently
implemented into the Monte-Carlo event generator \textsc{pegasus}.
The sea quark contribution is then incorporated in the former subprocess, 
while the small contribution of valence quarks can be taken into account
in the conventional QCD factorization (see\cite{Photon-our} for more details).
Numerically, we use the massless limit for light ($u$, $d$ and $s$) quarks, 
set the charm and beauty masses to $m_c = 1.4$~GeV and $m_b = 4.78$~GeV
and implement certain kinematical cut applied in the 
experimental analyses\footnote{All kinematic quantities are given in the laboratory frame with positive $OZ$ axis
directed as the proton beam.}:
$6 < E_T^\gamma < 15$~GeV, $-1.0 < \eta^\gamma < 2.4$, $0.1 < y < 0.7$
for H1 measurement\cite{gamma-H1} and $6 < E_T^\gamma < 15$~GeV, $-0.7 < \eta^\gamma < 0.9$ and $0.2 < y < 0.7$
for ZEUS\cite{gamma-ZEUS} one, where
$y$ is the fraction of the electron energy transferred to the photon (inelasticity). 
Both these data were obtained with the electron energy $E_e = 27.6$~GeV and proton energy $E_p = 920$~GeV. 
Results of our calculations for produced photon 
transverse energy and pseudorapidity distributions
are shown in Fig.~\ref{fig2}. 
One can see that our predictions (represented by the green histograms) 
are consistent with the latest HERA data within the experimental and theoretical uncertainties.
The latter are estimated in a traditional way, by varying the renormalization 
scale\footnote{In the CCFM-based approach, the factorization scale $\mu_F$ is related with the 
evolution variable and therefore should not be varied. See\cite{CCFM, JH2013} for more details.} 
around its default value $\mu_R = E_T^\gamma$ as $\mu_R \to \xi \mu_R$ with $\xi = 1/2$ or $2$.
In contrast with newly fitted LLM set, using the JH'2013 set 2 gluon
derived previously\cite{JH2013}
leads to systematic overestimation of the HERA data in most of the bins, that coincides with observations made earlier\cite{FL-our, Photon-our}. 
As it was argued\cite{LLM-2022}, better agreement achieved with the LLM distribution 
is a consequence of using a physically motivated expression~(\ref{eq:OurGluon1}) for the 
initial gluon density for subsequent CCFM evolution. Thus, our calculations
demonstrate that considered HERA data could help to distinguish the 
different approaches to evaluate the TMD gluon density in a proton.
Exact determination of the latter, of course, is important for experiments at future 
electron-proton or electron-ion colliders, such as NICA,
LHeC, FCC-eh, EiC and EiCC.

To summarize, we have refined phenomenological parameters of the TMD gluon density, 
which we obtained earlier (LLM). 
The analytical expression for initial distribution
provides a self-consistent simultaneous description of
HERA data on the proton structure function $F_2(x,Q^2)$, reduced cross section for the 
electron-proton deep inelastic scattering at low $Q^2$ 
and soft hadron production in $pp$ collisions at the LHC conditions.
It was evolved according to the CCFM equation to expand it to the whole kinematical region. 
Then our procedure was based on a fit to number of LHC and HERA data for 
processes sensitive to the gluon content of a proton.
The resulting fit quality ($\chi^2/d.o.f.= 1.773$) 
shows that the obtained gluon density does not contradict experimental data. 
We illustrate it additionally with latest HERA data on inclusive prompt photon photoproduction.
Our results together with the ones~\cite{LLM-F2} represent a self-consistent
approach for the TMD gluon density in a proton valid in a wide kinematical region. 
The updated LLM gluon density supersedes previous version and can be used in different phenomenological applications 
for $pp$, $p\bar p$ and $ep$ processes at modern and future colliders. It is available now in the 
\textsc{tmdlib} library and Monte-Carlo event generator \textsc{pegasus}.

{\sl Acknowledgements}. We thank S.P.~Baranov, A.A.~Prokhorov and
H.~Jung for their interest, important comments and remarks.
A.V.L. would like to thank School of Physics and Astronomy, Sun Yat-sen University (Zhuhai, China)
for warm hospitality.

\bibliography{F2}

\end{document}